\documentclass[twocolumn,twoside,slac_two]{revtex4}
\usepackage{epsfig}
\usepackage{graphicx}
\usepackage{fancyhdr}
\usepackage{subfigure}
\begin{document}

\title{The Ultra-High Energy Cosmic Ray Spectrum Measured by the
  Telescope Array's Middle Drum Detector}

%

\author{Douglas Rodriguez and Matt Wood for the Telescope Array Collaboration}
\affiliation{University of Utah, Salt Lake City, UT 84112, USA}
%

\begin{abstract}
The Telescope Array's Middle Drum fluorescence detector was
constructed using refurbished telescopes from the High Resolution
Fly's Eye (HiRes) experiment. As such, there is a direct comparison
between these two experiments' fluorescence energy spectra. An energy
spectrum has been calculated based on one year of collected data by the
Middle Drum site of Telescope Array and agrees well with the HiRes
monocular spectra. The quality of the Middle Drum results has also
been determined to show good agreement.

\end{abstract}

\maketitle

\thispagestyle{fancy}


\section{Telescope Array Middle Drum}
Telescope Array is a collaboration composed of institutions from the
U.S., Japan, Korea, Russia, and Belgium. The experiment consists of three
fluorescence telescope detectors overlooking 507
surface detectors arrayed over $\sim750~\rm{km^{2}}$. The northernmost
fluorescence detector, known as Middle Drum, consists of 14 telescopes
refurbished from the High Resolution Fly's Eye (HiRes)
experiment. These were arranged to view $\sim120^{\circ}$ in azimuthal
coverage and between $3^{\circ}$ and $31^{\circ}$ in elevation. Each
telescope unit uses sample-and-hold electronics with a $5.6 \mu s$
gate with floating tube-thresholds, allowing each photomultiplier tube
to have an individual firing rate of $\sim200$ Hz. Each telescope
camera consists of 256 photomultiplier tubes covered with an
ultra-violet transmissive filter. 

\subsection{Goals of Middle Drum}
The current goals of the Middle Drum spectral analysis are
three-fold. Since the Middle Drum detector has been operating
for over a year, the first goal of this analysis is to determine the
flux of particles observed during that first year using the same
analysis techniques used to produce the HiRes-1 monocular
spectrum. Secondly, since the telescope units are composed of the same
equipment used at HiRes, a direct comparison between Middle Drum and
HiRes can be performed. This results in a direct link in
the energy scale between these two experiments. Finally, by comparing
events observed by Middle Drum and any of the other detectors
(including other fluorescence detectors in either stereo or tandem
mode as well as the ground array), the energy scale across the entire
Telescope Array experiment can then be pinned down. This analysis will then
produce a bridge between the HiRes and the Telescope Array experiments.

\section{Exposure}
\begin{figure}[t]
 \centering
    \includegraphics[width=65mm]{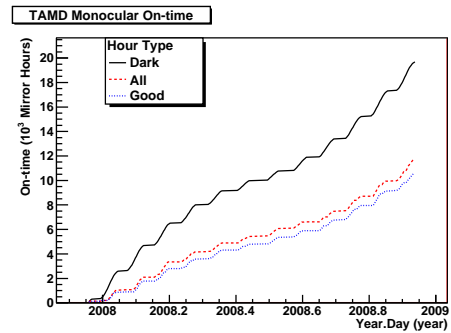}
    \caption{Middle Drum data collection time.} 
    \label{fig:ISVHECRI2010_53_1}
\end{figure}

The Telescope Array Middle Drum detector was refurbished and deployed
between November, 2006 and October, 2007. The first data was collected on
December 16, 2007 and the first year of collected data ended on
December 8, 2008. During this period, data was collected only on
nights that had at least three hours of full-dark: no sunlight and no
moonlight. Of this ``dark time'', many hours were lost due to poor
weather conditions in which the telescope bay doors were unable to be
opened. The final collected time (see figure
\ref{fig:ISVHECRI2010_53_1}) consisted of $\sim835$ hours ($\sim60\%$
of available time) of which $\sim754$ hours ($\sim90\%$ of collected
time) were considered good, i.e. minimal cloud cover in the view of
the detector. This correlates to a $\sim9\%$ duty cycle. This turned
out to be $\sim1/5$ the running time of HiRes-1. 

\begin{figure}[t]
  \centering
  \includegraphics[width=65mm]{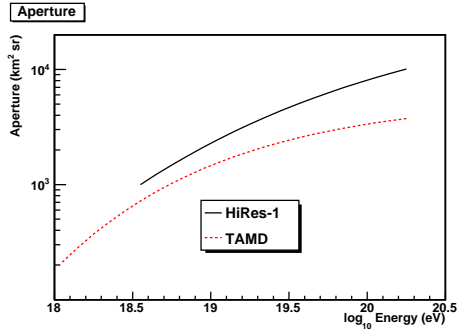}
  \caption{Middle Drum aperture compared to HiRes-1 aperture.} 
  \label{fig:ISVHECRI2010_53_2}
\end{figure}

The aperture of the Middle Drum detector is determined through Monte
Carlo simulations. 
The aperture of the Middle Drum detector has been calculated
to be roughly half that of the HiRes-1 detector (see figure
\ref{fig:ISVHECRI2010_53_2}). The combination of running time and
aperture is defined as the exposure of the detector. The exposure of
Middle Drum is then $\sim1/10$ the exposure of HiRes-1.

\section{The Middle Drum $1^{st}$ Year Spectrum}
\begin{figure}[t]
  \centering
  \includegraphics[width=65mm]{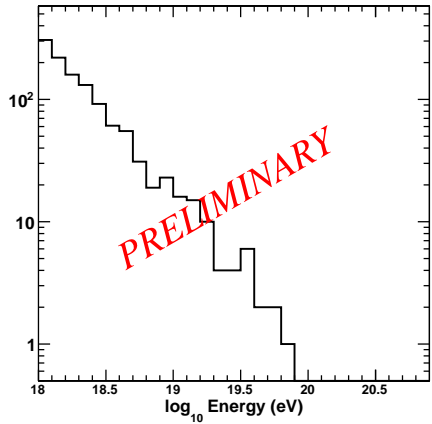}
  \caption{Middle Drum number of events observed.} 
  \label{fig:ISVHECRI2010_53_3}
\end{figure}

After one year of collecting data, 1156 events were observed with
energy above $10^{18.0}$ eV (see figure
\ref{fig:ISVHECRI2010_53_3}). There have been no observed events above
$10^{20.0}$ eV. Counting the number of events in each tenth-decadal
energy bin, it has been observed that there is $\sim1/10$ the number of
events observed by HiRes-1 for each year the Middle Drum detector
runs \cite{DCR_thesis}. As such, this is consistent with the exposure
for the first year of running.

\begin{figure}[t]
  \centering
  \includegraphics[width=65mm]{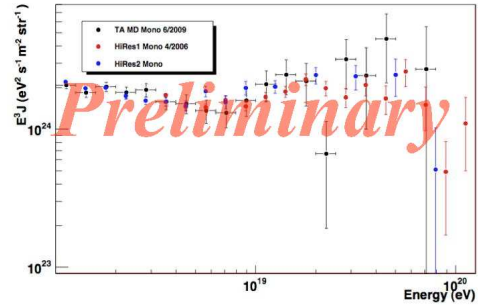}
  \caption{Middle Drum energy spectrum compared to HiRes.} 
  \label{fig:ISVHECRI2010_53_4}
\end{figure}

Combining the number of events and the exposure per energy bin, the
flux spectrum can be calculated. This results in a spectrum that is
consistent with that of the HiRes monocular spectrum (see figure
\ref{fig:ISVHECRI2010_53_4}) \cite{GZK_paper}.

\section{Monte Carlo}
The Monte Carlo was thrown with an isotropic distribution and sent
through the same processing cuts and reconstruction that was performed on the
data. The Monte Carlo was thrown with a Fly's Eye spectrum: a spectral
index of 3.2 below $10^{18.5}$ eV and 2.8 above that same energy,
between $10^{17.5}$ eV and $10^{21.0}$ eV. The
Monte Carlo spectral set was thrown without the GZK cutoff
\cite{PhysRevLett.16.748} \cite{SovPhysJETPL.4.78}. The HiRes-1 and
Telescope Array Middle Drum Monte Carlo uses the Gaisser-Hillas
parameterization to determine the number of charged particles at each
slant depth into the atmosphere. The number of photo-electrons
observed at each slant depth correlates to a certain energy deposited
into the atmosphere, and the integration of the energy deposited is
determined to be the energy of the initial cosmic ray. 

\subsection{Resolution}
\begin{figure}[ht]
  \centering
  \includegraphics[width=65mm]{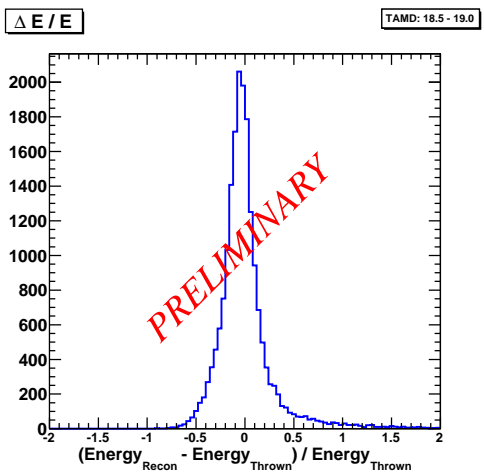}
  \caption{Middle Drum energy resolution for events with energy
    between $10^{18.5}$ and $10^{19.0}$ eV.} 
  \label{fig:ISVHECRI2010_53_5}
\end{figure}

\begin{figure}[ht]
  \centering
   \includegraphics[width=65mm]{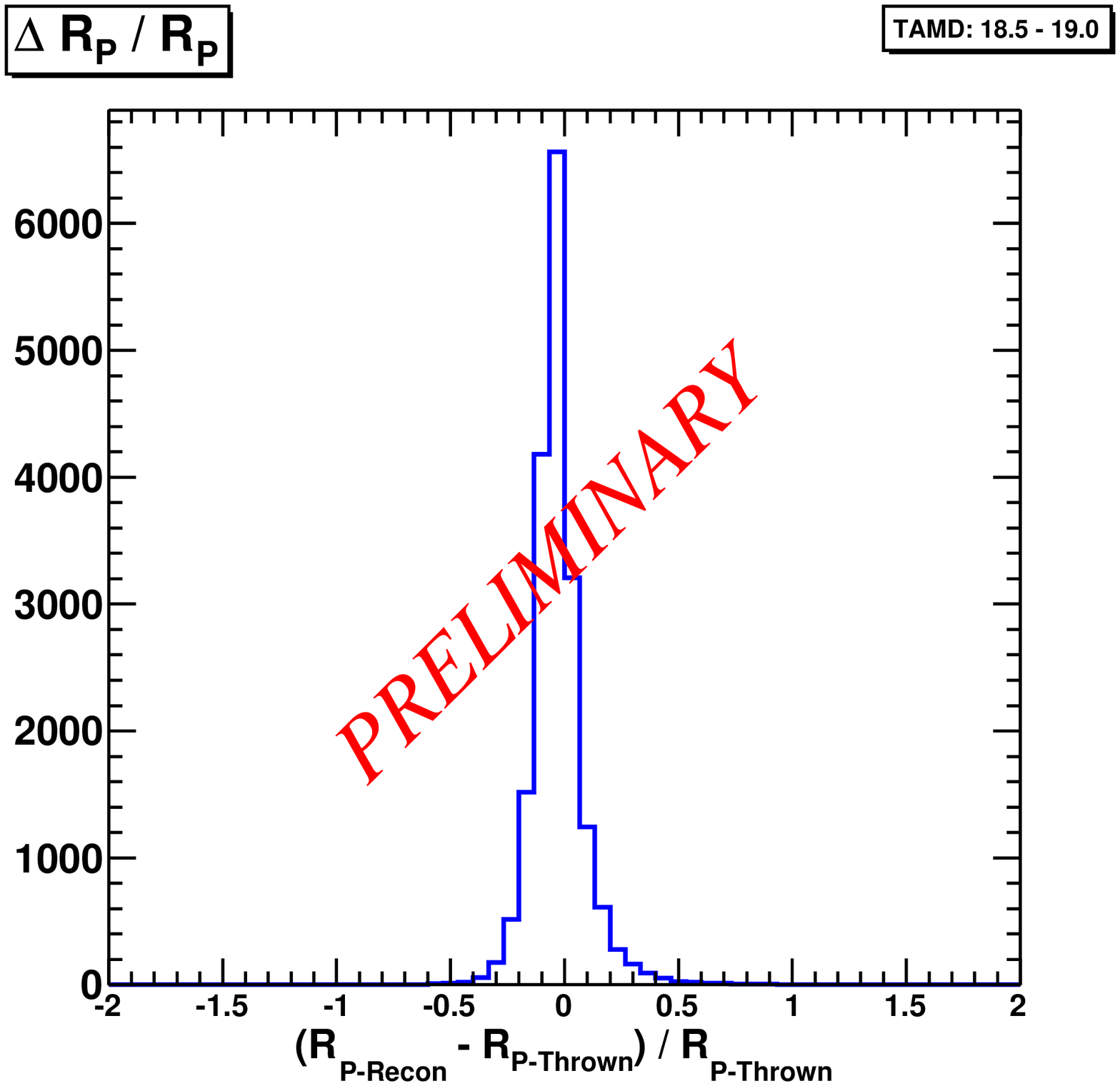}
 \caption{Middle Drum $R_{P}$ resolution for events with energy
    between $10^{18.5}$ and $10^{19.0}$ eV.} 
  \label{fig:ISVHECRI2010_53_6}
\end{figure}

\begin{figure}[ht]
  \centering
    \includegraphics[width=65mm]{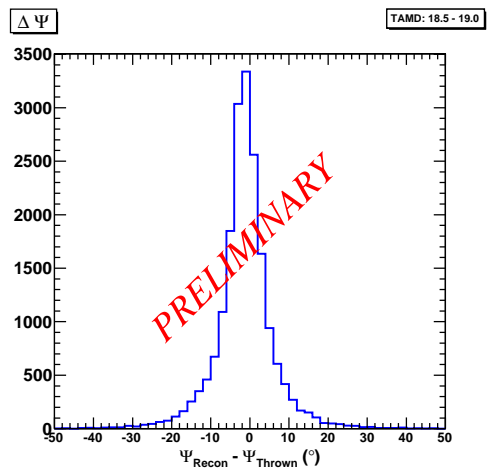}
 \caption{Middle Drum $\Psi$ resolution for events with energy
    between $10^{18.5}$ and $10^{19.0}$ eV.} 
  \label{fig:ISVHECRI2010_53_7}
\end{figure}

Resolution plots show how well the reconstruction programs
perform by determining the difference between the thrown and the
reconstructed energy and geometry in Monte Carlo simulations. The
three primary parameters that show the quality of the reconstruction
are the energy, the impact parameter, $R_{P}$, and the in-plane
angle, $\Psi$. These are determined for four energy ranges to show
trends in the reconstruction: $10^{17.5-18.0}$ eV, $10^{18.0-18.5}$
eV, $10^{18.5-19.0}$ eV, and $> 10^{19.0}$ eV. The lowest energy range
is below the spectrum limit and is not considered in the analysis and
the mid-range ($10^{18.5-19.0}$ eV) shows a good representation of
what is happening (see figures \ref{fig:ISVHECRI2010_53_5},
\ref{fig:ISVHECRI2010_53_6}, and \ref{fig:ISVHECRI2010_53_7}). 

\subsection{Data-MC Comparison}
\begin{figure}[ht]
  \centering
   \includegraphics[width=65mm]{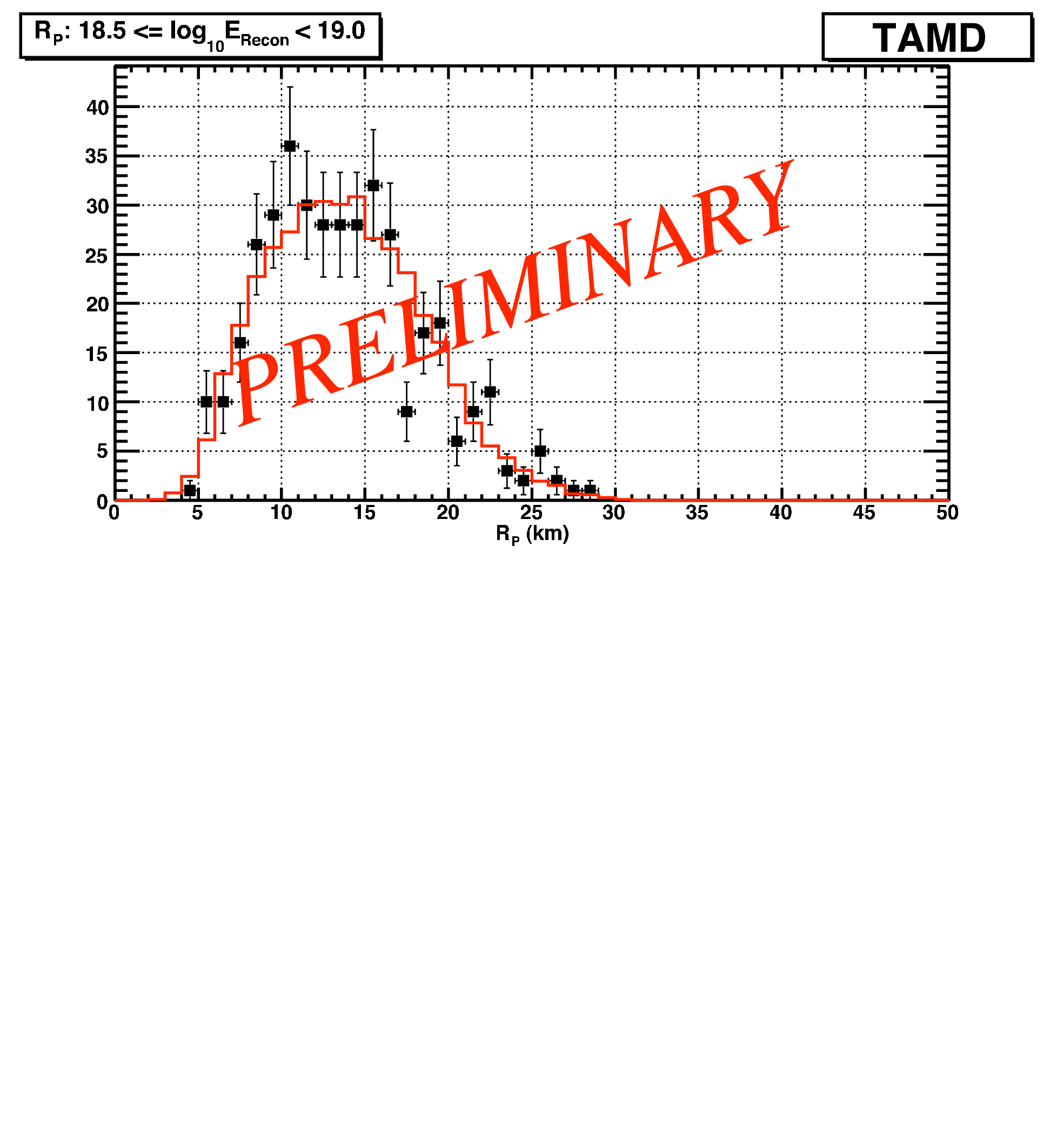}
 \caption{Middle Drum $R_{P}$ Data-Monte Carlo comparison for events with energy
    between $10^{18.5}$ and $10^{19.0}$ eV.} 
  \label{fig:ISVHECRI2010_53_8}
\end{figure}

\begin{figure}[ht]
  \centering
   \includegraphics[width=65mm]{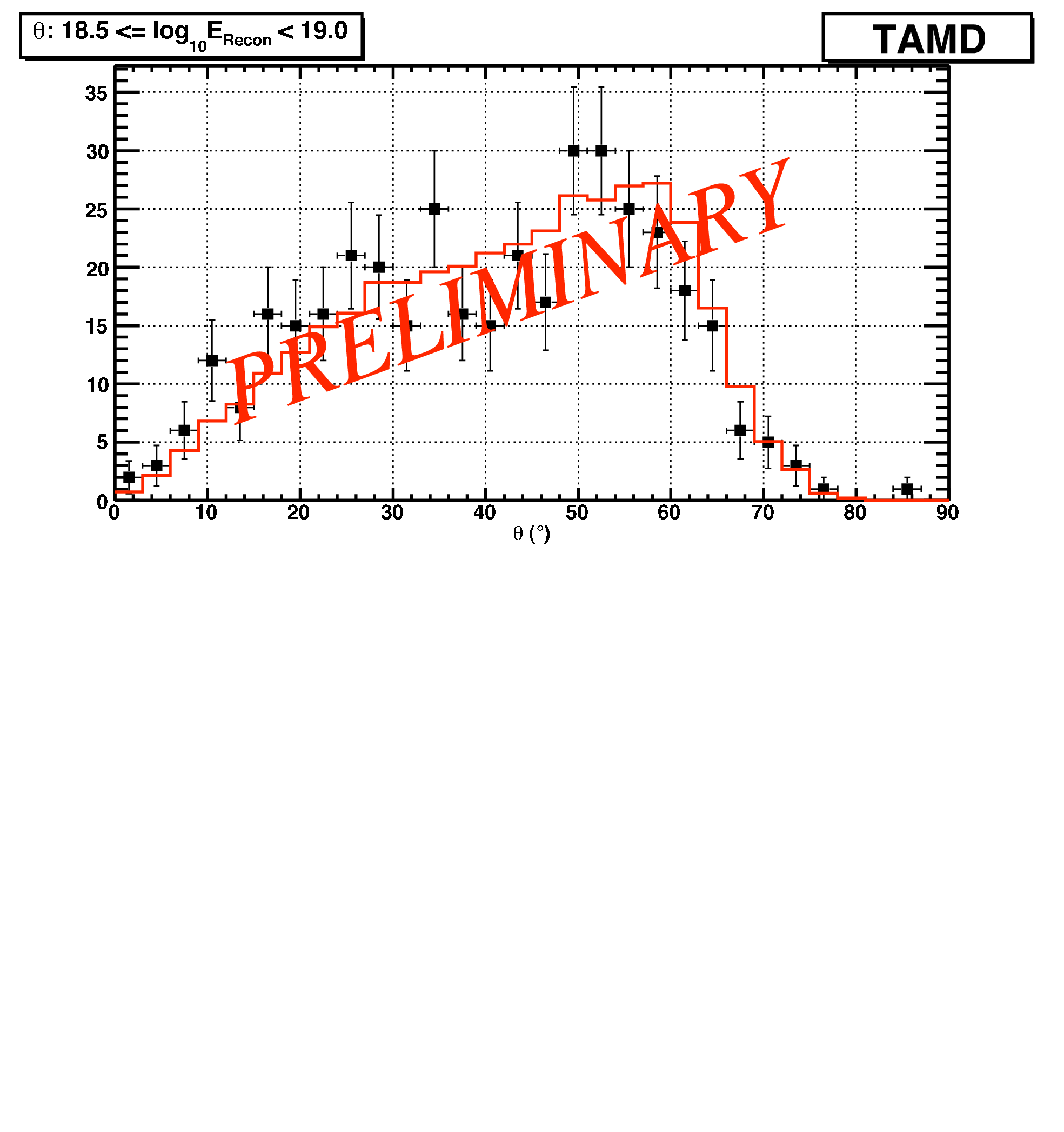}
 \caption{Middle Drum $\Psi$ Data-Monte Carlo comparison for events with energy
    between $10^{18.5}$ and $10^{19.0}$ eV.} 
  \label{fig:ISVHECRI2010_53_9}
\end{figure}

The aperture is solely based on the Monte Carlo simulation of real
events and how they are observed by our detector. In order to show how
well we can rely upon our aperture, data-Monte Carlo comparisons are
made. The same four energy ranges used in the resolution are again
compared here. However, the two most reasonable reconstruction
parameters to compare here are the impact parameter, $R_{P}$, and the
zenith angle, $\theta$. Again, the lowest energy range is not
considered since it is below the range of the spectrum and the
mid-range shows a reasonable trend (see figures \ref{fig:ISVHECRI2010_53_8} and
\ref{fig:ISVHECRI2010_53_9}). Due to the small statistics in the
actual data the error bars on the data points are quite
large. Nevertheless, it can be seen that the Monte Carlo histogram
lines overlay the data points very well.

\section{Conclusions}
The first year of the Telescope Array Middle Drum detector has been
analyzed using a monocular mode. The energy resolution of the Monte
Carlo simulations shows good agreement between what was thrown and
what was reconstructed. The data-Monte Carlo comparisons show good
agreement between simulated and real data. The calculated Middle Drum
energy spectrum is shown to be within statistical error of the spectra
produced by the HiRes monocular analysis.

\bigskip 
\begin{acknowledgments}
The Telescope Array experiment is supported by the Ministry of
Education, Culture, Sports, Science and Technology-Japan through
Kakenhi grants on priority area (431) “Highest Energy Cosmic Rays”,
basic research awards 18204020(A), 18403004(B) and 20340057(B); by the
U.S. National Science Foundation awards PHY-0307098, PHY-0601915,
PHY-0703893, PHY-0758342, and PHY-0848320 (Utah) and PHY-0649681
(Rutgers); by the Korea Research Foundation (KRF-2007-341-C00020); by
the Korean Science and Engineering Foundation (KOSEF,
R01-2007-000-21088-0); by the National Research Foundation of Korea
(NRF, 2010-0028071); by the Russian Academy of Sciences, RFBR grants
07-02-00820a and 09-07-00388a (INR), the FNRS contract 1.2.335.08,
IISN and Belgian Science Policy under IUAP VI/11 (ULB). The
foundations of Dr. Ezekiel R. and Edna Wattis Dumke, Willard L. Eccles
and the George S. and Dolores Dore Eccles all helped with generous
donations. The State of Utah supported the project through its
Economic Development Board, and the University of Utah through the
Office of the Vice President for Research. The experimental site became
available through the cooperation of the Utah School and Institutional
Trust Lands Administration (SITLA), U.S. Bureau of Land Management and
the U.S. Air Force. We also wish to thank the people and the officials
of Millard County, Utah, for their steadfast and warm supports. We
gratefully acknowledge the contributions from the technical staffs of
our home institutions and the University of Utah Center for High
Performance Computing (CHPC). 
\end{acknowledgments}

\bigskip 

\end{document}